\documentclass{aa}
\usepackage{graphicx}
\usepackage{natbib}
\usepackage{aas_macros}

\bibpunct{(}{)}{;}{a}{}{,}

\newcommand{\msun}{\mbox{${\rm M}_{\sun}$}}

\newcommand{\rsun}{\mbox{${\rm R}_{\sun}$}}

\begin{document}

\title{The formation of black hole low-mass X-ray binaries: through
  case B or case C mass transfer?}

\author{G. Nelemans and E.P.J. van den Heuvel\inst{1}}
\offprints{G. Nelemans, GijsN@astro.uva.nl}

\institute{Astronomical Institute ``Anton Pannekoek'', 
        Kruislaan 403, NL-1098 SJ Amsterdam, the Netherlands 
}

\date{accepted 3 July 2001} 

\titlerunning{Formation of black hole low-mass X-ray binaries}
\authorrunning{Nelemans \& van den Heuvel}

\abstract{The formation of low-mass X-ray binaries containing a rather
  massive ($M \ga 7 \msun$) black hole is problematic because in most
  recent stellar evolutionary calculations the immediate progenitors
  of these black holes (Wolf-Rayet stars) lose so much mass via their
  stellar wind that their final masses are well below the observed
  black hole masses.  We discuss the recently proposed solution that
  these binaries are formed through case C mass transfer (i.e. mass
  transfer after core helium burning is completed), avoiding a long
  Wolf-Rayet phase and thus significant mass loss. We show that only
  some of the currently available models for the evolution of massive
  stars allow this formation channel. We also investigate the effect
  of the downward revised Wolf-Rayet mass-loss rate as is suggested by
  observations, and conclude that in that case Wolf-Rayet stars end
  their lives with significantly higher masses than previously found
  and may be able to form a black holes.
}

\maketitle

\section{Introduction}

In low-mass X-ray binaries a neutron star or a black hole accretes
from a low-mass ($M \la 1 \msun$) companion.  A scenario to form such
stars begins with a relatively wide binary of a massive star and a
low-mass companion. When the massive star becomes a giant, mass
transfer is unstable and a common-envelope forms in which the
companion spirals down towards the core of the giant, leaving a close
binary consisting of the helium core of the giant and the low-mass
companion \citep{heu83}.  The helium star explodes in a supernova and
depending on the (core) mass of the helium star, a neutron star or
black hole is formed. With the discovery of A0620-00
\citep{esw+75,epp+75} and the determination of the mass function of
3.18 \citep{mr86}, the existence of the class of black hole low-mass
X-ray binaries was established. Currently we know 6 to 8 such systems
depending on the membership criteria \citep{cha98,bjc+98}. An
evolutionary scenario for these objects is given in \citet{khp87}.

To make a black hole, the initial mass of the primary must exceed a
critical value, which currently is believed to be around 20 \msun
\citep{fry99}. However, large mass-loss rates for massive stars and
Wolf-Rayet stars have been inferred from observations
\citep[e.g.][]{dnv88} and are found from the comparison of Wolf-Rayet
models with these observations \citep{lan89b}. Applying these rates to
evolutionary calculations resulted in the conclusion that even massive
single stars might end their evolution as relatively low-mass objects
when they explode \citep{ssm+92,mms+94,wlw95} and are thus unable to
produce the observed black holes \citep[see also][]{kal99}. For
massive stars in close binaries, which lose their hydrogen envelopes
due to mass transfer early in their evolution the situation is even
worse; the most recent calculations predict masses of helium stars as
they explode as low as 3 \msun, almost independent of their initial
mass \citep{wl99}.

In this article we first discuss the formation of black hole low-mass
X-ray binaries through case C evolution as suggested by \citet{blb99}
and \citet{wl99}: mass transfer starting after core-helium burning has
been completed \citep{kw67}. In this case a long-duration Wolf-Rayet
phase in which the star loses a lot of mass is avoided
(Sect.~\ref{caseC}).  Then we discuss the most recently observed
mass-loss rates for Wolf-Rayet stars and the implication of lower
mass-loss rates on the final helium-star masses of exploding stars in
binaries (Sect.~\ref{caseB}). At the end we discuss uncertainties and
possible alternatives for the formation of black hole low-mass X-ray
binaries (Sect.~\ref{discussion}) and end with our conclusions
(Sect.~\ref{conclusion}).

\section{Case C mass transfer}\label{caseC}

It has been suggested that case C mass transfer could be invoked to
avoid a long-duration Wolf-Rayet phase in the evolution of the massive
star, in order that this star does not lose too much mass and still is
able to form a massive black hole \citep{blb99,wl99}.

The occurrence of case C mass transfer depends on the radius evolution
of massive stars. For supergiants the radius of the star is not very
well defined, since the outer layers of the giant envelope are
extremely dilute. However, the best we can do is use the calculated
values of the radii of giants. We also neglect the interaction between
the wind of the massive star and the companion which may influence the
separation of the two stars.

We calculate the initial separation with which a binary should start
in order to undergo case C mass transfer as follows \citep[see
also][]{pve97}. The separation at the moment the Roche-lobe overflow
(RLOF) starts is given by
\begin{equation}
a_{\rm RLOF} =  \frac{R}{r_{\rm L}}
\end{equation}
where $R$ is the radius of the star and $r_{\rm L}$ is the
dimensionless Roche-lobe radius (the ratio of the Roche-lobe radius
and the binary separation). We use the \citet{egg83} equation for
$r_{\rm L}$. For mass ratios between 10 and 50, the value of $r_{\rm
  L}$ is between about 0.6 and 0.7. During the evolution the star
loses mass and the separation increases according to
\begin{equation}
a^{\prime} = a \, \frac{M}{M^{\prime}},
\end{equation}
where $M$ denotes the total mass of the binary.  So to start
Roche-lobe overflow at time $t$ when the star has a radius $R(t)$,
the
initial separation is given by
\begin{eqnarray}
a_{\rm i} & = & a_{\rm RLOF} (t) \,  \frac{M(t)}{M_{\rm i}} \nonumber
\\
    & = & \frac{R(t)}{r_{\rm L}}  \,  \frac{M(t)}{M_{\rm i}}.
\end{eqnarray}

\begin{figure}
  \resizebox{\columnwidth}{!}{\includegraphics[angle=-90]{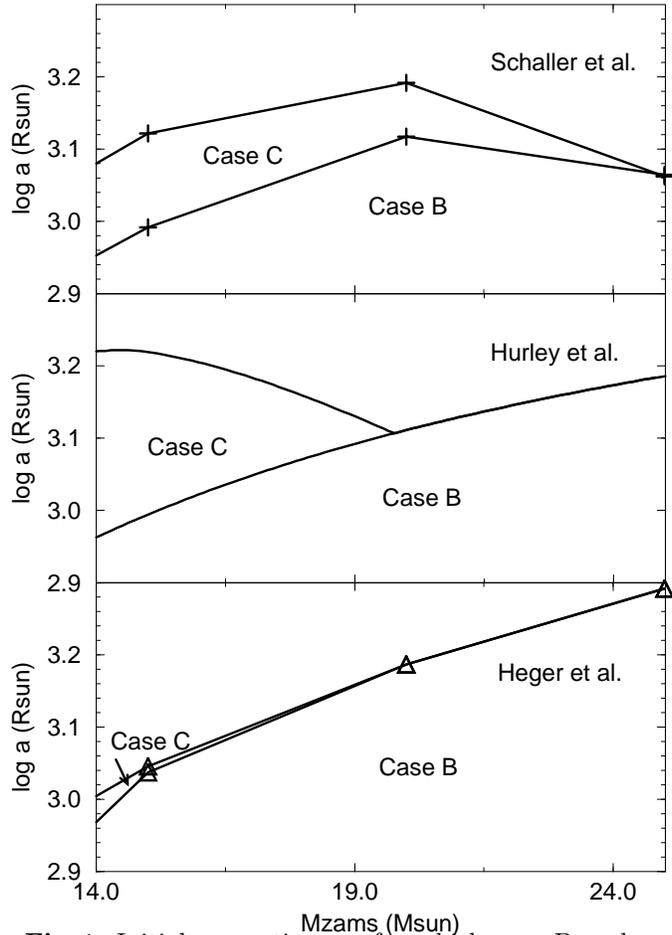}}
\caption[]{Initial separations  $a_{\rm i}$ for which case B and case
  C mass transfer occur as function of ZAMS mass, for a 1 \msun\ 
  companion.  \textbf{Top} for the \citet{ssm+92} models,
  \textbf{middle} for the \citet{hpt00} models, \textbf{bottom} for
  the \citet{hlw00} models.  }
\label{caseBC}
\end{figure}

We now compute the separations at which massive stars fill their Roche
lobes as function of initial mass and initial separation and determine
whether the mass transfer is case B or case C. In Fig.~\ref{caseBC}
(top) we show this for the evolutionary calculations of \citet[see
also Fig. 4 of \citealt{kw98}]{ssm+92}. For a star of initially 15
\msun case C mass transfer occurs for initial separation between 1000
and 1320 \rsun. For a 20 \msun star, these limits are 1300 and 1550
\rsun. For a 25 \msun star case C is not possible anymore. The two
other panels in Fig.~\ref{caseBC} show the same, but for the stellar
evolution models of \citet[middle]{hpt00} and \citet[bottom]{hlw00}.
For these models case C is not possible for stars more massive than
around 19 \msun. A recent estimate of the number of black hole
low-mass X-ray binaries that can form through the narrow case C
interval of the Schaller models shows that even such a narrow interval
might be enough to explain the whole Galactic population
\citep{blt01}.

\textsl{We conclude that since case C evolution depends strongly on
  the radius evolution of massive stars which is very uncertain, it
  seems possible but is not certain whether black hole low-mass X-ray
  binaries can be formed in this way.}

\section{Case B mass transfer}\label{caseB}

A different way to avoid too much mass loss may be the fact that
observed mass loss rates (which are the basis for the mass-loss rates
used in the evolutionary calculations) are revised downward
\citep{hk98,nl00}, which may make it possible to prevent helium stars
in binaries to lose so much mass they no longer can become black
holes. In a recent paper with drastically lower mass-loss rates,
derived from one particularly well-studied object and extrapolated,
final masses over 20 \msun for the most massive helium stars are found
\citep{che01}. As shown by \citet{kal99} the helium stars that were
the progenitors of the black holes in binaries cannot have lost more
than half of their initial mass. This includes both mass loss in the
stellar wind and in the supernova explosion.

\begin{figure}
\resizebox{\columnwidth}{!}{\includegraphics[angle=-90]{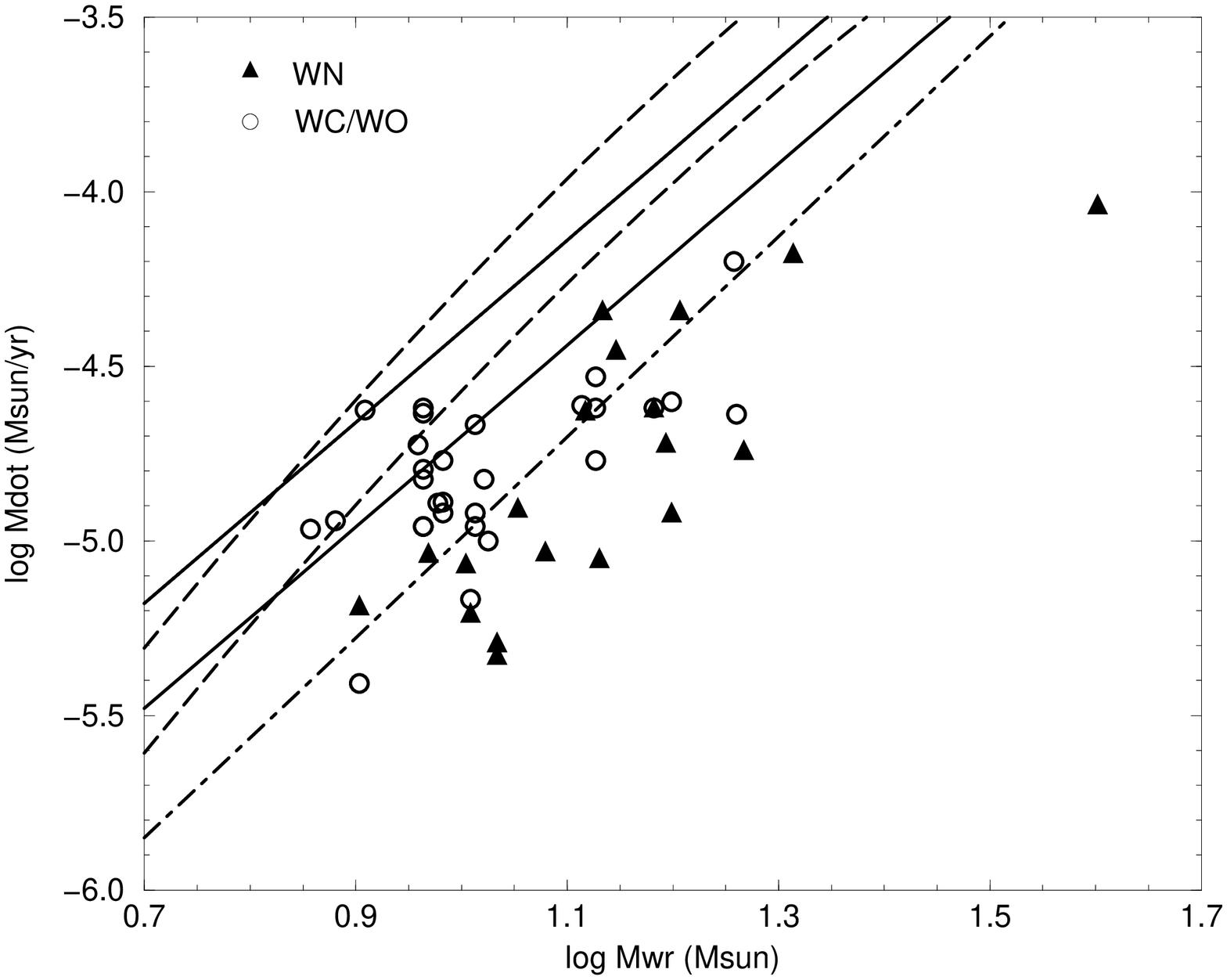}}
\caption[]{Mass-loss rates for Wolf-Rayet stars as observed
  \citep[triangles for WN and circles for WC/WO stars,
  from][where we excluded the hydrogen rich Wolf-Rayet
    stars]{nl00} and various relations used for evolutionary
  calculations. The solid lines are the relations assumed by
  \citet{wlw95} for WC (upper) and WN (lower) stars.  The dashed lines
  are the ones used by \citet{wl99}, where we converted the mass loss
  -- luminosity relations to mass loss - mass relations using the mass
  - luminosity relation of \citet{lan89a}. The upper dashed line is
  for their standard case, the lower for their reduced mass loss case.
  The dash-dotted line is a rectangular least square fit to all points
  (see also the text and Eq.~(\ref{mdot_GN})).
\label{Mdot}
}
\end{figure}

A recent compilation of observed mass-loss rates for Wolf-Rayet stars
is made by \citet{nl00}. In Fig.~\ref{Mdot} we show these inferred
mass-loss rates for WN and WC/WO stars (excluding the hydrogen rich
Wolf-Rayet stars). We overplotted mass-loss rates for WN and WC stars
as used by \citet{wlw95} as the solid lines. The mass-loss rates used
recently by \citet{wl99} are shown as the dashed lines, where we used
the luminosity -- mass relation as given by \citet{lan89a} to convert
the mass loss -- luminosity relation used by these authors, to a mass
loss -- mass relation. The top dashed line is their standard case, the
bottom a reduced mass-loss rate, which they used to account for the
lower observed mass-loss rates.

The most recently determined mass-loss rates thus suggest that the
rates used by \citet{wl99} are still too high. We will investigate the
effect of using a lower mass-loss rate law, which is shown in the
figure as the dash-dotted line and is given by
\begin{equation}\label{mdot_GN}
\dot{M} = -1.38 \, \times \, 10^{-8} \, M^{2.87}
\end{equation}
and is obtained by a `rectangular least square fit' \citep{lan89b} to
the data (i.e. minimising the rectangular distances to the line,
rather than the vertical distances). The fit is different from the one
obtained by \citet{nl00} because we excluded the hydrogen rich
Wolf-Rayet stars.

For a mass-loss rate of the form
\begin{equation}
\dot{M} = -k \, M^{\alpha}
\end{equation}
the final helium stars mass $M_{\rm f}$ can be computed from the
initial mass $M_{\rm i}$ and the helium star lifetime ($\tau$) from
\begin{equation}\label{mf_general}
  M_{\rm f} = \left[M_{\rm i}^{1 - \alpha} + (\alpha - 1) \, k \,
  \tau
  \right]^{1/(1 - \alpha)}.
\end{equation}

\begin{figure}
\begin{center}
\resizebox{\columnwidth}{!}{\includegraphics[angle=-90]{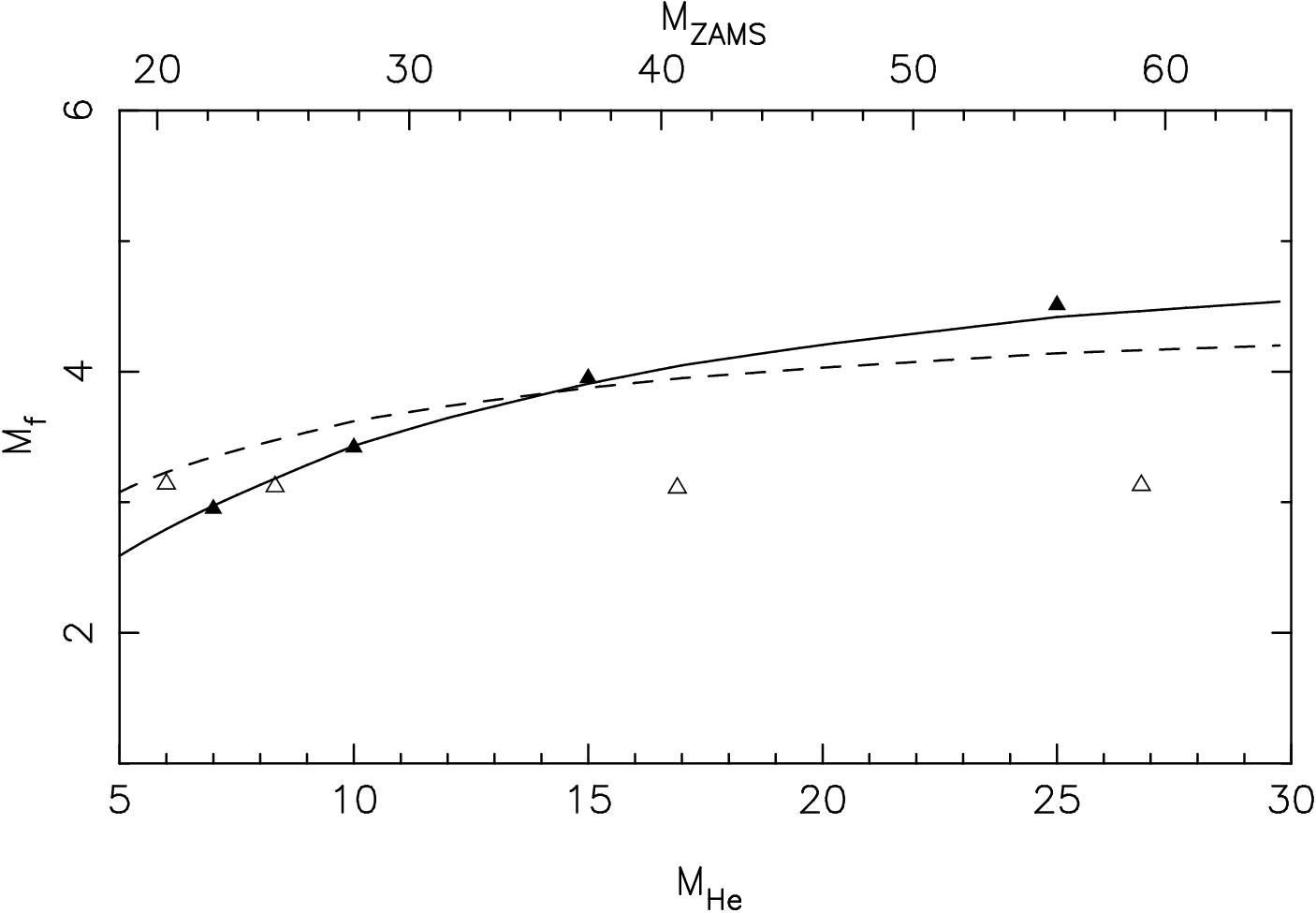}}
\caption[]{Final helium star masses as function of the initial helium
  star mass with the mass-loss rates according to \citet[solid line,
  some of their results are potted as solid triangles]{wlw95} and
  \citet[dashed line]{wl99} assuming a helium star lifetime as given
  by \citet{wlw95}. A selection of their results is plotted as the
  open triangles. The numbers at the top give an estimate of the ZAMS
  mass of the progenitor of the helium star.}
\label{hestar_low}
\end{center}
\end{figure}

As a check of our calculations we show in
  Fig.~\ref{hestar_low} the final masses that we obtain using the top
  two lines (dashed and solid) from Fig.~\ref{Mdot} and the helium
  star lifetimes as given by \citet{wlw95} and compare these with the
  results obtained with the same mass-loss rates by \citet{wlw95} and
  \citet{wl99}. For the dashed line the final mass is obtained by
  numerical integration of the mass evolution. The final masses do not
  completely agree with the masses obtained by \citet{wl99}, probably
  because these high mass-loss rates lead to even longer lifetimes.

\begin{figure}
\begin{center}
\resizebox{\columnwidth}{!}{\includegraphics[angle=-90]{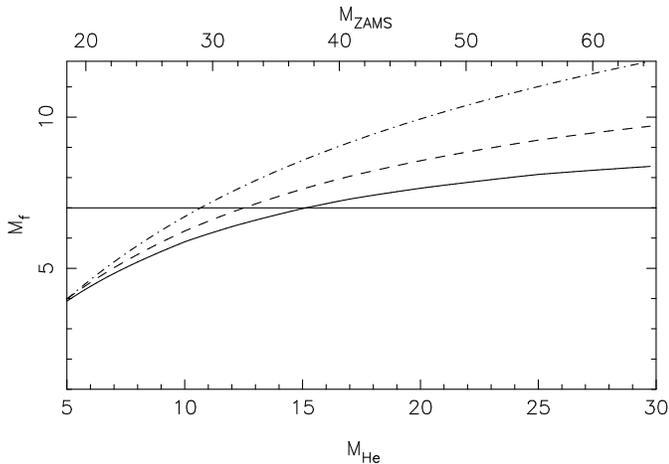}}
\caption[]{Final helium star masses as function of initial helium
  star mass with the mass-loss rates given by Eq.~(\ref{mdot_GN})
  using a helium-star lifetime as given by \citet[solid line]{wlw95}
  and the given by \citet[dash-dotted line]{pcw+91} and one with a
  helium-star lifetime halfway in-between these two (dashed
  line). ZAMS masses for the helium star progenitors are indicated at
  the top.}
\label{hestar_high}
\end{center}
\end{figure}

We now calculated the final masses for the revised mass-loss rate
given by Eq.~(\ref{mdot_GN}), which yields
\begin{equation}
M_{\rm f} = \left[M_{\rm i}^{-1.87} + 2.6 \, \times 10^{-8} \,
  \tau \right]^{-1/1.87}.
\end{equation}
In Fig.~\ref{hestar_high} we show these masses for the helium star
lifetimes from \citet[solid line]{wlw95}. The lifetime of the helium
star depends on the assumed mass-loss rate because mass-losing helium
stars become less massive, thus less luminous and can live longer.
For example the helium star lifetimes as given by \citet{wlw95} are
substantially longer than the ones collected by \citet{pcw+91} for
models without mass loss. We thus expect the lifetimes for the helium
stars with reduced mass-loss rates to be shorter. In
Fig.~\ref{hestar_high} we also plotted the final helium star masses
assuming a lifetime which is halfway in-between the lifetimes
given by \citet{wlw95} and \citet[dashed line]{pcw+91} and the one
given by \citet[dash-dotted line]{pcw+91}.

The horizontal line is at 7 \msun, the typical observed mass of the 
black holes in the low-mass X-ray binaries.  The limiting ZAMS mass
for which the final helium star mass exceeds 7 \msun with the masses
loss rate used here is $\sim30$ -- $37 \msun$.

\textsl{We thus conclude that with revised mass-loss rates helium
  stars end their lives with significantly higher masses than
  previously found and may be able to form black holes even after
  case
  B mass transfer.}

\section{Discussion}\label{discussion}

The analysis in Sect.~\ref{caseC} neglects the influence of the wind
of the massive star on the companion star. The companion moves through
the wind and already feels friction, which counteracts the widening of
the orbit due to the stellar wind. However, even for a wind mass-loss
rate of $10^{-3} \msun$~yr$^{-1}$, the density in the wind at the
companion is almost four orders of magnitude lower than the density at
the edge of the giant for a giant with radius of 1000 \rsun and a
binary separation of 1600 \rsun.

The whole argument presented in Sect.~\ref{caseB} is based on the
observed mass-loss rates. However, it should be noted that \emph{all}
mass-loss rates proposed for Wolf-Rayet stars and used in evolutionary
calculations are based on the observed rates. The valid question still
remains what the uncertainty is in the observed mass-loss rates and in
the inferred stellar masses and how this could influence our main
conclusion.

The mass-loss rates as determined by \citet{nl00} are the most
accurate, but still suffer from the general problem that not all
quantities (mass, mass-loss rate and luminosity) can be determined
independently. They therefore use the mass -- luminosity relation of
\citet{sm92} to obtain the final mass estimates from the luminosity.
Using a different mass -- luminosity relation may change the resulting
mass/mass-loss rate combinations.

Taking the masses and mass-loss rates as plotted in Fig.~\ref{Mdot},
one would not say that there is a unique mass-loss rate -- mass
relation, as is expected on theoretical grounds \citep{lan89a}.  The
scatter is larger than the quoted uncertainty in the observations.
This either points to underestimates of the errors in the
observations, to variability or to additional physical processes,
which were not taken into account in the calculations by
\citet{lan89a} and can change the mass-loss rate for a given
Wolf-Rayet star mass. One could think of rotation, magnetic fields or
maybe the evolutionary history.

In the last respect it might be that stars in binaries that lose their
hydrogen envelopes by mass transfer evolve differently from stars that
lose their envelopes due to their own stellar winds (which possibly is
enhanced by a companion). The question which stars actually form black
holes and which neutron stars is considerably more complex than the
question of the final mass of helium stars \citep[e.g.][]{fry99}. In
particular the evolution of the core is important. As long as the
collapse of the core are not understood this question will remain
unanswered.

Finally, it should be noted that to form a black hole low-mass X-ray
binary the companion must survive the common-envelope phase. The
outcome of the common envelope depends on the binding energy and
density structure of the giants envelope, which are quite different
for giants that undergo case B and case C mass transfer. It could for
instance be that that all binaries that undergo case B mass transfer
to a low-mass companion will completely merge. That would mean that we
\emph{need} the small allowed initial separation range for case C.

\section{Conclusion}\label{conclusion}

We calculated the possible initial separations for which case C mass
transfer is likely to occur for binaries containing a massive
  star and a low-mass star, using different stellar evolution models.
We find that case C mass transfer becomes impossible for
primaries more massive than around 19 \msun for the models of
\citet{hlw00} and \citet{hpt00} and more massive than around 25 \msun
for the models by \citet{ssm+92}. For such binaries either
  case B mass transfer occurs, or no mass transfer at all.  Unless
the current models for massive stars underestimate the radius
expansion after the end of core helium burning the chances for forming
black holes in binaries through case C mass transfer are therefore
limited.

We also investigated the influence of the assumed mass-loss rate on
the final mass of helium stars in binaries and conclude that with a
downward revised mass-loss rate as suggested by the observations
\citep[e.g.][]{nl00} helium stars end their lives with significantly
higher masses than previously found and may be able to form black
holes even after case B mass transfer for primaries more massive than
$\sim30$ -- $40 \msun$.

\begin{acknowledgements}
  We thank Jasinta Dewi and Norbert Langer for helpful discussion and
  trial calculations of helium star evolution and Alexander Heger for
  providing details of his evolutionary calculations. This work was
  supported by NWO Spinoza grant 08-0 to E.~P.~J.~van den Heuvel.
\end{acknowledgements}

\bibliography{journals,binaries}
\bibliographystyle{apj}

\end{document}